\documentclass[aps, twocolumn, pra, showpacs, superscriptaddress]{revtex4}
\usepackage{graphicx}
\usepackage{amsmath}
\usepackage{amsfonts}
\usepackage{amssymb}
\usepackage{epsfig}

\begin{document}

\title{Fidelity susceptibility and long-range correlation in the
Kitaev honeycomb model}

\author{Shuo Yang}
\affiliation{Department of Physics and ITP, The Chinese University of Hong
Kong, Hong Kong, China} \affiliation{Institute of Theoretical Physics, Chinese
Academy of Sciences, Beijing, 100080, China}

\author{Shi-Jian Gu}  \email{sjgu@phy.cuhk.edu.hk}
\affiliation{Department of Physics and ITP, The Chinese University of Hong
Kong, Hong Kong, China}

\author{Chang-Pu Sun}
\affiliation{Institute of Theoretical Physics, Chinese Academy of Sciences,
Beijing, 100080, China}

\author{Hai-Qing Lin}
\affiliation{Department of Physics and ITP, The Chinese University of Hong
Kong, Hong Kong, China}

\date{\today }

\begin{abstract}
We study exactly both the ground-state fidelity susceptibility and
bond-bond correlation function in the Kitaev honeycomb model. Our
results show that the fidelity susceptibility can be used to
identify the topological phase transition from a gapped A phase with
Abelian anyon excitations to a gapless B phase with non-Abelian
anyon excitations. We also find that the bond-bond correlation
function decays exponentially in the gapped phase, but algebraically
in the gapless phase. For the former case, the correlation length is
found to be $1/\xi=2\sinh^{-1}[\sqrt{2J_z -1}/(1-J_z)]$, which
diverges around the critical point $J_z=(1/2)^+$.
\end{abstract}





\maketitle

\section{Introduction}

Quite recently, a great deal of effort
\cite{HTQuan2006,Zanardi06,Pzanardi0606130,Buonsante1,PZanardi032109,WXG,
PZanardi0701061,WLYou07,HQZhou07,LCVenuti07,SChen07,SJGu072,MFYang07,
NPaunkovic07,zhq,AHamma07} has been devoted to the role of fidelity,
a concept borrowed from quantum information theory \cite{Nielsen1},
in quantum phase transitions(QPTs)\cite{Sachdev}. The motivation is
quite obvious. Since the fidelity is a measure of similarity between
two states, the change of the ground state structure around the
quantum critical point should result in a dramatic change in the
fidelity across the critical point. Such a fascinating prospect has
been demonstrated in many correlated systems. For example, in the
one-dimensional XY model, the fidelity shows a narrow trough at the
phase transition point \cite{Zanardi06}. Similar properties were
also found in fermionic \cite{Pzanardi0606130} and bosonic systems
\cite{Buonsante1}. The advantage of the fidelity is that, since the
fidelity is a space geometrical quantity, no a priori knowledge of
the order parameter and symmetry-breaking is required in studies of
QPTs.

Nevertheless, the properties of the fidelity are mainly determined
by its leading term \cite{PZanardi0701061,WLYou07}, i.e., its second
derivative with respect to the driving parameter (or the so-called
fidelity susceptibility \cite{WLYou07}). According to the standard
perturbation method, it has been shown that the fidelity
susceptibility actually is equivalent to the structure factor
(fluctuation) of the driving term in the Hamiltonian \cite{WLYou07}.
For example, if we focus on the thermal phase transitions and choose
the temperature as the driving parameter, the fidelity
susceptibility, extracted from the mixed state fidelity between two
thermal states\cite{WXG}, is simply the specific
heat\cite{PZanardi0701061,WLYou07}. From this point of view, the
fidelity approach to QPTs seems still to be within the framework of
the correlation functions approach, which is intrinsically related
to the local order parameter.

However, some systems cannot be described in a framework built on
the local order parameter. This might be due to the absence of
preexisting symmetry in the Hamiltonian, such as topological phase
transitions \cite{wen-book} and Kosterlitz-Thouless phase
transitions \cite{JMKosterlitz73}. For the latter, since the
transition is of infinite-order, it has already been pointed out
that the fidelity might fail to identify the phase transition point
\cite{WLYou07,SChen07}. Therefore, it is an interesting issue to
address the role of fidelity in studying the topological phase
transition.

The Kitaev honeycomb model was first introduced by Kitaev in search
of topological order and anyonic statistics. The model is associated
with a system of 1/2 spins which are located at the vertices of a
honeycomb lattice. Each spin interacts with three nearest neighbor
spins through three types of bonds, called ``$x$($y,z$)-bonds"
depending on their direction. The model Hamiltonian \cite{Kitaev} is
as follows:
\begin{eqnarray}
H &=& -J_{x}\sum_{x\text{-bonds}}\sigma _{j}^{x}\sigma _{k}^{x}-J_{y}\sum_{y
\text{-bonds}}\sigma _{j}^{y}\sigma _{k}^{y}-J_{z}\sum_{z\text{-bonds} }\sigma
_{j}^{z}\sigma _{k}^{z},\nonumber
\\
&=& -J_x H_x - J_y H_y - J_z H_z. \label{eq:Hamiltonian}
\end{eqnarray}
where $j, k$ denote the two ends of the corresponding bond, and
$J_a, \sigma^a(a=x,y,z)$ are dimensionless coupling constants and
Pauli matrices respectively. Such a model is rather artificial.
However, its potential application in topological quantum
computation has made it a focus of research in recent years
\cite{Kitaev,Wen,preskill9,pachos,Sarma,XYFeng07,CHD,YuYue,YS,spKou,model3D,Dusuel}.

The ground state of the Kitaev honeycomb model consists of two
phases, i.e., a gapped A phase with Abelian anyon excitations and a
gapless B phase with non-Abelian anyon excitations. The transition
has been studied by various approaches. For example, it has been
shown that a kind of long range order exists in the dual space
\cite{XYFeng07}, such that basic concepts of Landau's theory of
continuous phase transitions might still be applied. In real space,
however, the spin-spin correlation functions vanishes rapidly with
increasing distance between two spins. Therefore, the transition
between the two phases is believed to be of topological type due to
the absence of a local order parameter in real space \cite{Kitaev}.

In this work, we \emph{firstly} try to investigate the topological
QPT occurring in the ground state of the Kitaev honeycomb model in
terms of the fidelity susceptibility. We find that the fidelity
susceptibility can be used to identify the topological phase
transition from a gapped phase with Abelian anyon excitations to
gapless phase with non-Abelian anyon excitations. Various scaling
and critical exponents of the fidelity susceptibility around the
critical points are obtained through a standard finite-size scaling
analysis. \emph{These observations from the fidelity approach are a
little surprising}. Our earlier thought was that the fidelity
susceptibility, which is a kind of structure factor obtained by a
combination of correlation functions, can hardly be related to the
topological phase transition, since the latter cannot be described
by the correlation functions of local operators. So our
\emph{second} motivation following from the first one is to study
the dominant correlation function appearing in the definition of the
fidelity susceptibility, i.e., the bond-bond correlation function.
We find that the correlation function decays algebraically in the
gapless phase, but exponentially in the gapped phase. For the
latter, the correlation length takes the form
$1/\xi=2\sinh^{-1}[\sqrt{2J_z -1}/(1-J_z)]$ along a given evolution
line. Therefore, the divergence of the correlation length around the
critical point $J_z=(1/2)^+$ is also a signature of the QPT.

We organize our work as follows. In Sec. \ref{sec:gsfs}, we
introduce briefly the definition of the fidelity susceptibility in
the Hamiltonian parameter space, then we diagonalize the Hamiltonian
based on Kitaev's approaches and obtain the explicit forms of the
Riemann metric tensor, from which the fidelity susceptibility along
any direction can be obtained. The critical and scaling behaviors of
the fidelity susceptibility are also studied numerically. In Sec.
\ref{sec:lrc}, we explicitly calculate the bond-bond correlation
functions in both phases. Its long range behavior and the
correlation length in the gapped phase are studied both analytically
and numerically. Sec. \ref{sec:sum} includes a brief summary.

\section{Fidelity susceptibility in the ground state}
\label{sec:gsfs}

To study the fidelity susceptibility, we notice that the structure
of the parameter space of the Hamiltonian (\ref{eq:Hamiltonian}) is
three dimensional. In this space, we can always let the ground state
of the Hamiltonian evolves along a certain path in the parameter
space, i.e.,
\begin{eqnarray}
J_a=J_a(\lambda),
\end{eqnarray}
where $\lambda$ is a kind of driving parameter along the evolution
line. We then extend the definition of fidelity to this arbitrary
line in high-dimensional space. Following Ref. \cite{Zanardi06}, the
fidelity is defined as the overlap between two ground states
\begin{eqnarray}
F=|\langle \Psi_0(\lambda) |\Psi_0(\lambda+\delta\lambda)\rangle|,
\end{eqnarray}
where $\delta\lambda$ is the magnitude of a small displacement along the
tangent direction at $\lambda$. Then the fidelity susceptibility along this
line can be calculated as
\begin{eqnarray}
\chi_F = \lim_{\delta\lambda\rightarrow 0}\frac{-2\ln F_i}{\delta\lambda^2} =
\sum_{a b} g_{a b} n^a n^b, \label{eq:deffsmetric}
\end{eqnarray}
where $n^a=\partial J_a /\partial \lambda$ denotes the tangent unit
vector at the given point, and $g_{ab}$ is the Riemann metric tensor
introduced by Zanardi, Giorda, and Cozzini\cite{PZanardi0701061}.
For the present model, we have
\begin{eqnarray}
g_{ab}=\sum_n \frac{\langle \Psi_n(\lambda) |H_a|\Psi_0(\lambda) \rangle
\langle \Psi_0(\lambda) |H_b| \Psi_n(\lambda) \rangle}{(E_n -E_0)^2},
\label{eq:metric}
\end{eqnarray}
where $|\Psi_n(\lambda) \rangle$ is the eigenstate of the
Hamiltonian with energy $E_n$. Clearly, $g_{ab}$ does not depend on
the specific path along which the system evolves. However, once
$g_{ab}$ are obtained, the fidelity susceptibility is just a simple
combination of $g_{ab}$ together with a unit vector which defines
the direction of system evolution in the parameter space.

According to Kitaev \cite{Kitaev}, the Hamiltonian
(\ref{eq:Hamiltonian}) can be diagonalized exactly by introducing
Majorana fermion operators to represent the Pauli operators as
\begin{eqnarray}
\sigma ^{x}=\text{i}b^{x}c, \; \sigma ^{y}=\text{i}b^{y}c,\; \sigma
^{z}=\text{i}b^{z}c, \label{eq:spinmajorana}
\end{eqnarray}
where the Majorana operators satisfy $A ^{2}=1$, $AB =-BA $ for $A
,B \in \left\{ b^{x},b^{y},b^{z},c\right\} $ and $A \neq B $, and
also $b^{x}b^{y}b^{z}c\left\vert \psi \right\rangle =\left\vert \psi
\right\rangle $ to ensure the commutation relations of spin
operators. Then the Hamiltonian can be written as
\begin{eqnarray}
H= \frac{\text{i}}{2} \sum_{j,k}\widehat{u}_{jk}J_{a_{jk}}c_{j}c_{k}.
\end{eqnarray}
Since the operators $ \widehat{u}_{jk}=\text{i}b_{j}^{a _{jk}}b_{k}^{a _{jk}}$
satisfy $\left[
\widehat{u}_{jk},H\right] =0$, $\left[ \widehat{u}_{jk},\widehat{u}_{ml}%
\right] =0$, and $\widehat{u}_{jk}^{2}=1$, they can be regarded as
generators of the $Z_{2}$ symmetry group. Therefore, the whole
Hilbert space can be decomposed into common eigenspaces of
$\widehat{u}_{jk}$, each subspace is characterized by a group of
$u_{jk}=\pm 1$. The spin model is transformed to a quadratic
Majorana fermionic Hamiltonian
\begin{eqnarray}
H=\frac{\text{i}}{2} \sum_{j,k}u_{jk}J_{a _{jk}}c_{j}c_{k}.
\end{eqnarray}
Here we restrict ourselves to only the vortex free subspace with
translational invariants, i.e., all $u_{jk}=1$. After Fourier
transformation, we get the Hamiltonian of a unit cell in the
momentum representation \cite{Kitaev},
\begin{equation}
H=\sum_{\textbf{q}}\left(\begin{array}{c}a_{-\textbf{q},1}
\\a_{-\textbf{q},2}\end{array}
\right) ^{\mathrm{T}}\left(\begin{array}{cc}0 & \text{i}f\left( \textbf{q}\right)  \\
-\text{i}f\left( \textbf{q}\right) ^{\ast } & 0\end{array} \right) \left(
\begin{array}{c}a_{\textbf{q},1} \\a_{\textbf{q},2}\end{array}\right) ,
\end{equation}
where $\textbf{q}=(q_x, q_y)$,
\begin{equation}
a_{\textbf{q},\gamma
}=\frac{1}{\sqrt{2L^{2}}}\sum_{\mathbf{r}}e^{-\text{i}\mathbf{q}\cdot
\mathbf{r}}c_{\mathbf{r},\gamma },\label{eq:aqcq}
\end{equation}
$\mathbf{r}$ refers to the coordinate of a unit cell, $\gamma $ to a
position type inside the cell, and
\begin{eqnarray}
f\left( \textbf{q}\right)  &=&\epsilon _{\textbf{q}}+\text{i}\Delta _{\textbf{q}},  \nonumber \\
\epsilon _{\textbf{q}} &=&J_{x}\cos q_{x}+J_{y}\cos q_{y}+J_{z},  \nonumber \\
\Delta _{\textbf{q}} &=&J_{x}\sin q_{x}+J_{y}\sin q_{y}.
\end{eqnarray}
Here, we set $L$ to be an odd integer, then the system size is
$N=2L^{2}$. The momenta take the values
\begin{equation}
q_{x\left( y\right) }=\frac{2n\pi }{L},n=-\frac{L-1}{2},\cdots
,\frac{L-1}{2} .
\end{equation}
The above Hamiltonian can be rewritten using fermionic operators as
\begin{equation}
H=\sum_{\textbf{q}}\sqrt{\epsilon _{\textbf{q}}^{2}+\Delta
_{\textbf{q}}^{2}}\left( C_{\textbf{q},1}^{\dag
}C_{\textbf{q},1}-C_{\textbf{q},2}^{\dag }C_{\textbf{q},2}\right) .
\end{equation}
Therefore, we have the ground state
\begin{eqnarray}
\left\vert \Psi _{0}\right\rangle  &=&\prod_{\textbf{q}}C_{\textbf{q},2}^{\dag
}\left\vert
0\right\rangle   \nonumber \\
&=&\prod_{\textbf{q}}\frac{1}{\sqrt{2}}\left( \frac{\sqrt{\epsilon
_{\textbf{q}}^{2}+\Delta _{\textbf{q}}^{2}}}{\Delta
_{\textbf{q}}+\text{i}\epsilon
_{\textbf{q}}}a_{-\textbf{q},1}+a_{-\textbf{q},2}\right) \left\vert
0\right\rangle ,\label{eq:ground}
\end{eqnarray}
with the ground state energy
\begin{equation}
E_{0}=-\sum_{\textbf{q}}\sqrt{\epsilon _{\textbf{q}}^{2}+\Delta
_{\textbf{q}}^{2}}.
\end{equation}

\begin{figure}
\includegraphics[bb=30 370 550 765, width=8.5 cm, clip] {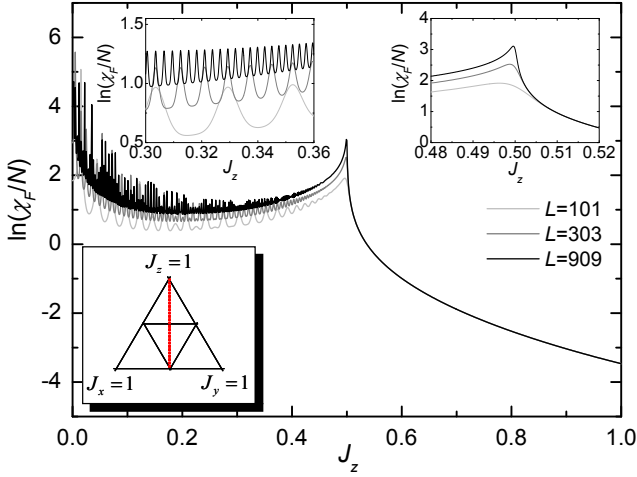}
\caption{(Color online) Fidelity susceptibility as a function of
$J_z$ along the dashed line shown in the triangle for various system
sizes $L=101, 303, 909$. Both upper insets correspond to enlarged
pictures of two small portions.} \label{figure_fs}
\end{figure}

The fidelity of the two ground states at $\lambda$ and $\lambda'$
can be obtained as
\begin{eqnarray}
F^2 &=&\prod_{\textbf{q}}\frac{1}{2}\left( 1+\frac{\Delta _{\textbf{q}}\Delta
_{\textbf{q}}^{\prime }+\epsilon _{\textbf{q}}\epsilon _{\textbf{q}}^{\prime
}}{E_{\textbf{q}}E_{\textbf{q}}^{\prime }}
\right), \nonumber \\
&=&\prod_{\textbf{q}}\cos ^{2}\left( \theta _{\textbf{q}}-\theta
_{\textbf{q}}^{\prime }\right).
\end{eqnarray}
with
\begin{eqnarray}
\cos \left( 2\theta _{\textbf{q}}\right)  &=&\frac{\epsilon
_{\textbf{q}}}{E_{\textbf{q}}},\sin \left(
2\theta _{\textbf{q}}\right) =\frac{\Delta _{\textbf{q}}}{E_{\textbf{q}}}, \nonumber \\
\cos \left( 2\theta _{\textbf{q}}^{\prime }\right)  &=&\frac{\epsilon _{\textbf{q}}^{\prime }}{%
E_{\textbf{q}}^{\prime }},\sin \left( 2\theta _{\textbf{q}}^{\prime }\right)
=\frac{\Delta _{\textbf{q}}^{\prime }}{E_{\textbf{q}}^{\prime }}.
\end{eqnarray}
The Riemann metric tensor can be expressed as
\begin{eqnarray}
g^{ab} =\sum_{\textbf{q}}\left( \frac{\partial \theta _{\textbf{q}}}{\partial J_{a}}%
\right) \left( \frac{\partial \theta _{\textbf{q}}}{\partial J_{b}}\right),
\end{eqnarray}
where
\begin{eqnarray}
\frac{\partial \left( 2\theta _{\textbf{q}}\right) }{\partial J_{x}} &=&\frac{%
J_{z}\sin q_{x}+J_{y}\sin \left( q_{x}-q_{y}\right) }{\epsilon
_{\textbf{q}}^{2}+\Delta _{\textbf{q}}^{2}}\cdot \frac{\Delta
_{\textbf{q}}}{\left\vert \Delta
_{\textbf{q}}\right\vert }, \nonumber \\
\frac{\partial \left( 2\theta _{\textbf{q}}\right) }{\partial J_{y}} &=&-\frac{%
J_{x}\sin \left( q_{x}-q_{y}\right) -J_{z}\sin q_{y}}{\epsilon
_{\textbf{q}}^{2}+\Delta _{\textbf{q}}^{2}}\cdot \frac{\Delta
_{\textbf{q}}}{\left\vert \Delta
_{\textbf{q}}\right\vert }, \nonumber \\
\frac{\partial \left( 2\theta _{\textbf{q}}\right) }{\partial J_{z}} &=&-\frac{%
J_{x}\sin q_{x}+J_{y}\sin q_{y}}{\epsilon _{\textbf{q}}^{2}+\Delta
_{\textbf{q}}^{2}}\cdot \frac{\Delta _{\textbf{q}}}{\left\vert \Delta
_{\textbf{q}}\right\vert }.
\end{eqnarray}
Clearly, with these equations, we can in principle calculate the fidelity
susceptibility along any direction in the parameter space according to Eq.
(\ref{eq:deffsmetric}). Here, we would like to point out that the same results
can be obtained from the generalized Jordan-Wigner transformation used firstly
by Feng, Zhang, and Xiang\cite{XYFeng07}.

\begin{figure}
\includegraphics[bb=3 355 540 772, width=8.5 cm, clip] {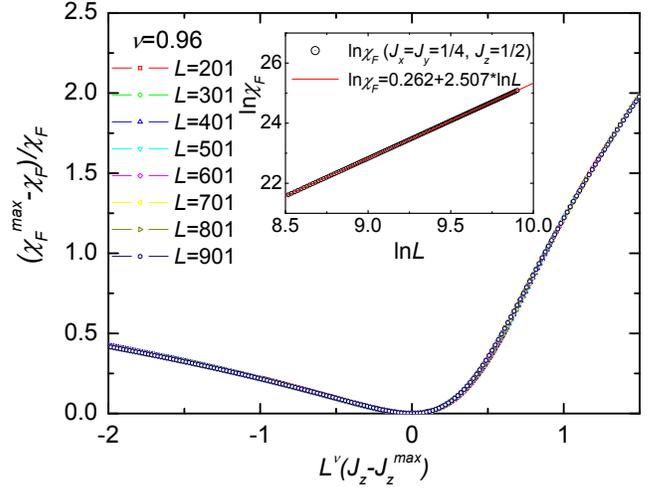}
\caption{ (Color online) Finite size scaling analysis for the case
of power-law divergence for system sizes $L=201, 301, \dots, 901$.
The fidelity susceptibility, considered as a function of system size
and driving parameter is a function of $L^\nu (J_z -J_z^{\rm max})$
only, and has the critical exponent $\nu=0.96$.}
\label{figure_scale}
\end{figure}

Following Kitaev \cite{Kitaev}, we restrict our studies to the plane
$J_x+J_y+J_z=1$ (see the large triangle in Fig. \ref{figure_fs}).
According to his results, the plane consists of two phases, i.e., a
gapped A phase with Abelian anyon excitations and a gapless B phase
with non-Abelian excitations. The two phases are separated by three
transition lines, i.e. $J_x=1/2$, $J_y=1/2$, and $J_z=1/2$ which
form a small triangle in the B phase.

Generally, we can define an arbitrary evolution line on the plane. Without loss
of generality, we first choose the line as $J_{x}=J_{y}$ (see the dashed line
in the triangle of Fig. \ref{figure_fs}). Then the fidelity susceptibility
along this line can be simplified as
\begin{eqnarray}
\chi _{F} = \frac{1}{16}\sum_{\textbf{q}}\left[ \frac{\sin q_{x}+\sin
q_{y}}{\epsilon _{\textbf{q}}^{2}+\Delta _{\textbf{q}}^{2}}\right] ^{2}.
\end{eqnarray}
The numerical results of different system sizes are shown in Fig.
\ref{figure_fs}. First of all, the fidelity susceptibility per site,
i.e. $\chi_F/N$ diverges quickly with increasing system size around
the critical point $J_z=1/2$. This property is similar to the
fidelity susceptibility in other systems, such as the
one-dimensional Ising chain \cite{Zanardi06} and the asymmetric
Hubbard model\cite{SJGu072}. Secondly, $\chi_F/N$ is an intensive
quantity in the A phase ($J_z>1/2$), while in the B phase, the
fidelity susceptibility also diverges with increasing system size.
Thirdly, the fidelity susceptibility shows many peaks in the B
phase, the number of peaks increases linearly with the system size
$L$ (see the left upper inset of Fig. \ref{figure_fs}). The
phenomena of fidelity susceptibility per site in the B phase have
not been found in other systems previously, to our knowledge, so
that they are rather impressive.

\begin{figure}
\includegraphics[bb=20 360 560 770, width=8.5 cm, clip] {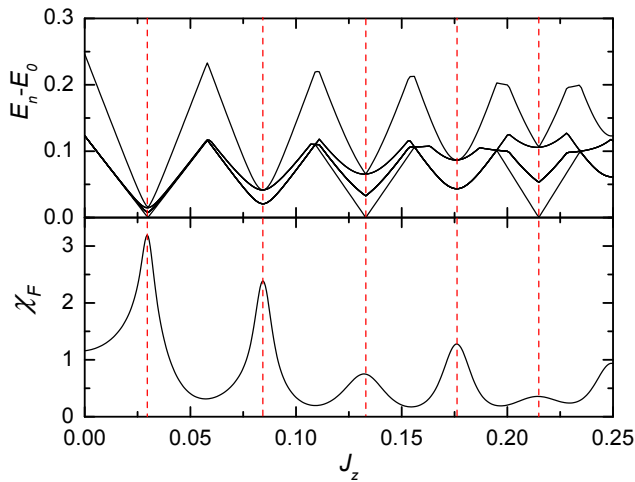}
\caption{(Color online) Fidelity susceptibility and a few low-lying
excitations as a function of $J_z$ in a small portion of the
evolution line for system size $L=51$.} \label{figure_fsen}
\end{figure}

To study the scaling behavior of the fidelity susceptibility around
the critical point, we perform a finite-size scaling analysis. Since
the fidelity susceptibility in the A phase is an intensive quantity,
the fidelity susceptibility in the thermodynamic limit, scales as
\cite{SJGu072}
\begin{eqnarray}
\frac{\chi_F}{N} \propto \frac{1}{|J_z -J_z^c|^\alpha}.
\end{eqnarray}
around $J_z^c=1/2$. Meanwhile, the maximum point of $\chi_F$ at
$J_z=J_z^{\rm max}$ for a finite sample behaves as
\begin{eqnarray}
\frac{\chi_F}{N} \propto L^\mu,
\end{eqnarray}
with $\mu=0.507\pm 0.0001$ (see the inset of Fig. \ref{figure_scale}).
According to the scaling ansatz, the rescaled fidelity susceptibility around
its maximum point at $J_z^{\rm max}$ is just a simple function of the rescaled
driving parameter, i.e.,
\begin{eqnarray}
\frac{\chi_F^{\rm max}-\chi_F}{\chi_F} =f[L^\nu (J_z - J_z^{\rm max})].
\end{eqnarray}
where $f(x)$ is a universal scaling function and does not depend on
the system size, and $\nu$ is the critical exponent. The function
$f(x)$ is shown in Fig. \ref{figure_scale}. Clearly, the rescaled
fidelity susceptibilities of various system sizes fall onto a single
line for a specific $\nu=0.96\pm 0.005$. Then the critical exponent
$\alpha$ can be obtained as
\begin{eqnarray}
\alpha=\frac{\mu}{\nu}=0.528\pm 0.001.
\end{eqnarray}

\begin{figure}
\includegraphics[bb=50 350 560 760, width=8 cm, clip] {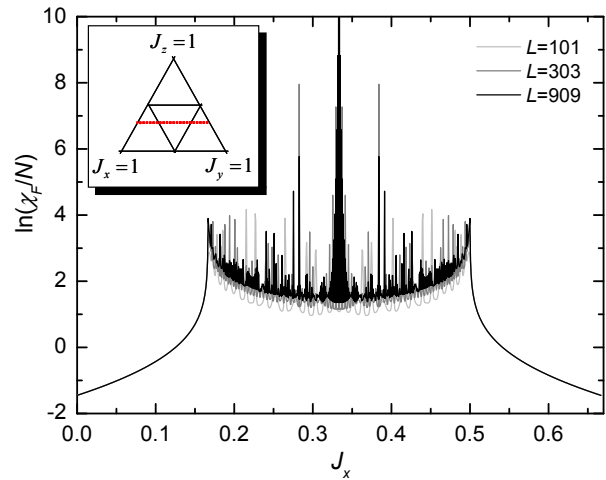}
\caption{(Color online) Fidelity susceptibility as a function of
$J_x=2/3-J_y$ along the dashed line shown in the triangle for
various system sizes $L=101, 303, 909$. } \label{figure_fs2}
\end{figure}

One of the most interesting observations is that a huge number of
peaks appear in the B phase. The scaling analysis shows that the
number of peaks is proportional to the system size. Physically, a
peak means that the ground state can not adiabatically evolve from
one side of the peak to the other side easily because the two ground
states have distinct features. From this point of view, the ground
state in the B phase might be stable to a adiabatic perturbation.
Moreover, the existence of many peaks can also be reflected by
reconstruction of the energy spectra. For this purpose, we choose a
small portion of the evolution line and plot both the fidelity
susceptibility and a few low-lying excitations in Fig.
\ref{figure_fsen}. Since the fidelity is inversely proportional to
the energy gap [Eq. (\ref{eq:metric})], the location of each peak
corresponds to a gap minimum.

Similarly, we can also choose the system evolution line as
$J_z=1/3$, the fidelity susceptibility then takes the form
\begin{eqnarray}
\chi _{F} = \frac{1}{36}\sum_{\textbf{q}}\left[ \frac{\left( \sin q_{x}-\sin
q_{y}\right) +2\sin \left( q_{x}-q_{y}\right) }{\epsilon
_{\textbf{q}}^{2}+\Delta _{\textbf{q}}^{2}}\right] ^{2}.
\end{eqnarray}
The numerical results for this case are shown in Fig.
\ref{figure_fs2}. The results are qualitatively similar to those of
previous cases. In the B phase, there still exist many peaks. Both
the number and the magnitude of the peaks increase with the system
size, while in the A phase, the fidelity susceptibility becomes an
intensive quantity.

\section{Long-range correlation and fidelity susceptibility}
\label{sec:lrc}

Follow You, {\it et al.} \cite{WLYou07}, the fidelity susceptibility
is a combination of correlation functions. Precisely, for a general
Hamiltonian
\begin{eqnarray}
H=H_0+\lambda H_I,
\end{eqnarray}
the fidelity susceptibility can be calculated as
\begin{eqnarray}
\chi_F=\int\tau\left[ \langle \Psi_0 |H_I(\tau) H_I(0)|\Psi_0\rangle
-\langle\Psi_0|H_I|\Psi_0\rangle^2\right]
d\tau\label{eq:fidelityfnal},
\end{eqnarray}
with $\tau$ being the imaginary time and $$
H_I(\tau)=e^{H(\lambda)\tau} H_I e^{-H(\lambda)\tau}.$$ Therefore,
the divergence of the fidelity susceptibility at the critical point
implies the existence of a long-range correlation function. Without
loss of generality, if we still restrict ourselves to the plane
$J_x+J_y+J_z=1$ and choose $J_z$ ($J_{x}=J_{y}$) as the driving
parameter, the bond-bond correlation function is defined as
\begin{eqnarray}
C\left( \mathbf{r}_{1},\mathbf{r}_{2}\right)  &=&\left\langle \sigma
_{ \mathbf{r}_{1},1}^{z}\sigma _{\mathbf{r}_{1},2}^{z}\sigma
_{\mathbf{r}
_{2},1}^{z}\sigma _{\mathbf{r}_{2},2}^{z}\right\rangle \nonumber \\
&&-\left\langle \sigma _{\mathbf{r}_{1},1}^{z}\sigma _{\mathbf{r}
_{1},2}^{z}\right\rangle \left\langle \sigma
_{\mathbf{r}_{2},1}^{z}\sigma _{ \mathbf{r}_{2},2}^{z}\right\rangle
\label{eq:correlationdef}.
\end{eqnarray}
Here the subscripts $\mathbf{r}_{1},1$ and $\mathbf{r}_{1},2$ denote
the two ends of the single $z$-bond at $\mathbf{r}_{1}$=($x, y$). In
the vortex-free case, through Eqs. (\ref{eq:spinmajorana}),
(\ref{eq:aqcq}), and (\ref{eq:ground}), the spin operators $\sigma
_{ \mathbf{r}_{1},1}^{z}\sigma _{\mathbf{r}_{1},2}^{z}$ can be
expressed in the form of fermion operators. So we finally get
\begin{eqnarray}
\left\langle \sigma _{\mathbf{r}_{1},1}^{z}\sigma _{\mathbf{r}
_{1},2}^{z}\right\rangle =\left\langle \sigma
_{\mathbf{r}_{2},1}^{z}\sigma _{\mathbf{r}_{2},2}^{z}\right\rangle
=\frac{1}{N}\sum_{\mathbf{q}}\frac{ \epsilon
_{\mathbf{q}}}{E_{\mathbf{q}}}
\end{eqnarray}
and
\begin{eqnarray}
&&\left\langle \Psi _{0}\right\vert \sigma _{\mathbf{r}_{1},1}^{z}\sigma _{%
\mathbf{r}_{1},2}^{z}\sigma _{\mathbf{r}_{2},1}^{z}\sigma _{\mathbf{r}%
_{2},2}^{z}\left\vert \Psi _{0}\right\rangle \nonumber \\
&=&\frac{1}{N^{2}}\sum_{\mathbf{q},\mathbf{q}^{\prime }}\left\{ \cos
\left[
\left( \mathbf{q}-\mathbf{q}^{\prime }\right) \left( \mathbf{r}_{1}-\mathbf{r%
}_{2}\right) \right] -1\right\} \nonumber \\
&&\times \frac{\left( \Delta _{\mathbf{q}}\Delta _{\mathbf{q}^{\prime }}-\epsilon _{%
\mathbf{q}}\epsilon _{\mathbf{q}^{\prime }}\right)}{ E_{\mathbf{q}}E_{\mathbf{%
q}^{\prime }}} \label{eq:Correlation4343}
\end{eqnarray}
with $\mathbf{q}\neq\mathbf{q}^{\prime }$ and
$\mathbf{r}_{1}\neq\mathbf{r}_{2}$. The same results can also be
obtained by using the Jordan-Wigner transformation method
\cite{XYFeng07,CHD}.

\begin{figure}
\includegraphics[bb=5 373 503 765, width=8.5 cm, clip]{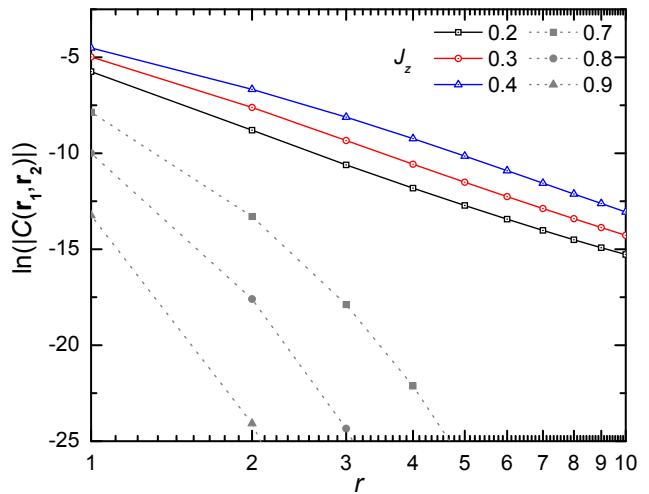}
\caption{(Color online) Bond-bond correlation function as a function
of distance $r$ for various $J_z$ and a finite sample of $L=100$,
where $\textbf{r}_1 -\textbf{r}_2=(r,r)$. Downward peaks in top
lines are due to zero-point crossing.} \label{figure_correlation}
\end{figure}

\begin{figure}
\includegraphics[bb=53 367 575 793, width=8.5 cm, clip] {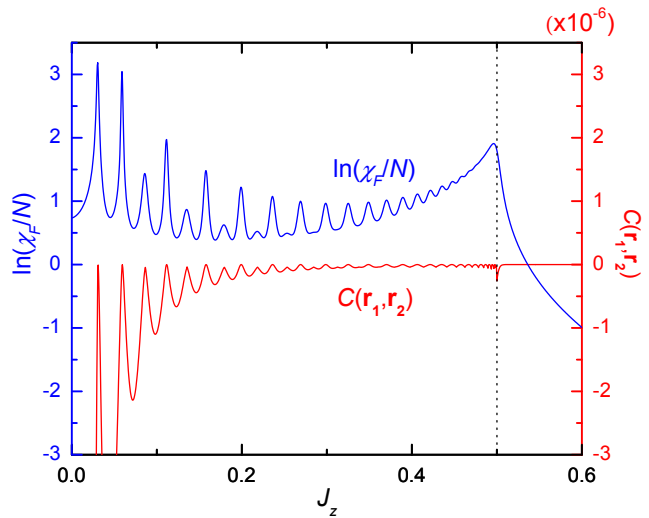}
\caption{(Color online) Fidelity susceptibility and the correlation
function at $\textbf{r}_1 -\textbf{r}_2=(L/2, L/2) $ as a function
of $J_z$ for a finite sample of $L=100$. } \label{figure_fscorr}
\end{figure}

We show the dependence of the correlation function Eq.
(\ref{eq:correlationdef}) on the distance for a finite sample of
$L=100$ in Fig. \ref{figure_correlation}. Obviously, the lines can
be divided into two groups. In the gapless phase ($J_z<1/2$), the
correlation function decays algebraically, while in the gapped phase
($J_z>1/2$), it decays exponentially. If $J_z<1/2$, the denominator
in Eq. (\ref{eq:Correlation4343}) has two zero points, which are of
order $1/N$ in the large $N$ limit. Their contribution causes the
summation to be finite in the thermodynamic limit. Then using the
stationary phase method, we can evaluate the exponents of the
correlation function at long distance to be 4, i.e.,
\begin{eqnarray}
C\left(\mathbf{r}_1,\mathbf{r}_2\right) \propto
\frac{1}{|\mathbf{r}_1-\mathbf{r}_2|^4}.
\end{eqnarray}
From Fig. \ref{figure_correlation}, the average slope of the top
three lines around $r=10$ is estimated to be $4.05$, which is
slightly different from $4$. Nevertheless, we would rather interpret
the difference as due to both finite size effects and numerical
error. On the other hand, if $J_z>1/2$, the phase is gapped and the
denominator in Eq. (\ref{eq:Correlation4343}) does not have zero
point on the real axis. Therefore, the whole summation is strongly
suppressed except for the case of small
$|\mathbf{r}_{1}-\mathbf{r}_{2}|$, whose range actually defines the
correlation length. In order to evaluate the correlation, we need to
extend the integrand (in the thermodynamic limit) in Eq.
(\ref{eq:Correlation4343}) to the whole complex plane, where we can
find two singular points. Using the steepest descent method, we can
evaluate the correlation length to be
\begin{eqnarray}
\frac{1}{\xi}=2\sinh^{-1}\frac{\sqrt{2J_z -1}}{1-J_z}.
\end{eqnarray}
Obviously, the correlation length becomes divergent as
$J_z\rightarrow 0.5^+$. This property can also be used to signal the
QPT occurring in the Kitaev honeycomb model in addition to the
fidelity and Chern number \cite{Kitaev}. The correlation length we
obtained is the same as that of the string operators \cite{CHD},
which, however, is a non-local operator.

Although it is not easy to calculate the fidelity susceptibility
from the correlation function directly due to the dynamic term in
Eq. (\ref{eq:fidelityfnal}), our conjecture is confirmed for the
present model. That is the divergence of the fidelity susceptibility
is related to the long-range correlations. Fig. \ref{figure_fscorr}
is illustrative. The correlation function at $\textbf{r}_1
-\textbf{r}_2=(L/2, L/2)$, in spite of its smallness, remains
nonzero in the region $J_z < 1/2$, but it vanishes in $J_z
> 1/2$. For the former, the oscillating structures of the two lines meet each
other.

\section{summary and discussion}
\label{sec:sum}

In summary, we have studied the critical behavior of the fidelity
susceptibility where a topological phase transition occurrs in the
honeycomb Kitaev model. Though no symmetry breaking exists and no
local order parameter in real space can be used to describe the
transition, the fidelity susceptibility definitely can indicate the
transition point. We found that the fidelity susceptibility per site
is an intensive quantity in the gapped phase, while in the gapless
phase, the huge number of peaks reflects frequent spectral
reconstruction along the evolution line. We also studied various
scaling and critical exponents of the fidelity susceptibility around
the critical points.

Based on the conclusions from the fidelity, we further studied the
bond-bond correlation function in both phases. We found that the
bond-bond correlation function, which plays a dominant role in the
expression for the fidelity susceptibility, decays exponentially in
the gapped phase, but algebraically in the gapless phase. The
critical exponents of the correlation function in both the gapless
and gapped phases are calculated numerical and analytically.
Therefore, in addition to the topological properties of the Kitaev
honeycomb model, say, the Chern number, we found that both the
fidelity susceptibility and the bond-bond correlation functions can
be used to witness the QPT in the model.

\emph{Note added}. After finishing this work, we noticed that a work
on the fidelity per site instead of the fidelity susceptibility in a
similar model appeared\cite{JHZhao0803}.

\begin{acknowledgements}
We thank Xiao-Gang Wen, Yu-Peng Wang, Guang-Ming Zhang, and Jun-Peng
Cao for helpful discussions. This work is supported by CUHK (Grant
No. A/C 2060344) and NSFC.
\end{acknowledgements}

\end{document}